%Paper: 9112044
%From: VENEZIA%CERNVM@pucc.PRINCETON.EDU
%Date: Tue, 17 Dec 91 12:41:00 SET

\magnification=1200
\hsize 15true cm \hoffset=0.5true cm
\vsize 23true cm
\baselineskip=15pt

\outer\def\beginsection#1\par{\medbreak\bigskip
      \message{#1}\leftline{\bf#1}\nobreak\medskip\vskip-\parskip
      \noindent}

\def \pa {\partial}
\def \ra {\rightarrow}
\def \pr {\prime}
\def \se {\prime \prime}
\def \ti {\tilde}

\def \a {\alpha}

\def \Ga {\Gamma}
\def \ga {\gamma}
\def \sg {\sigma}
\def \da {\delta}
\def \ep {\epsilon}
\def \r {\rho}
\def \om {\omega}
\def \Om {\Omega}
\def \noi {\noindent}

\def\sqr#1#2{{\vcenter{\hrule height.#2pt\hbox{\vrule width.#2pt
height#1pt \kern#1pt\vrule width.#2pt}\hrule height.#2pt}}}

\def\lsim{\mathrel{\rlap{\lower4pt\hbox{\hskip1pt$\sim$}}
    \raise1pt\hbox{$<$}}}         %less than or approx. symbol
\def\gsim{\mathrel{\rlap{\lower4pt\hbox{\hskip1pt$\sim$}}
    \raise1pt\hbox{$>$}}}         %greater than or approx. symbol

 \nopagenumbers

\null\vskip-.5cm
{\hfill\ CERN-TH.6321/91}\par
\line{\hfil DFTT-50/91}
{\hfill\ November 1991}
\vskip2cm
\centerline{\bf $O(d,d)$-COVARIANT STRING COSMOLOGY}

\vskip.5cm
\centerline{\bf M. Gasperini}
\centerline{\it Dipartimento di Fisica Teorica}

\centerline{\it and}
 \centerline{\it INFN, Sezione di Torino}
\vskip.5cm
\centerline{\bf  G. Veneziano }
\centerline{\it Theory Division, CERN, Geneva, Switzerland}
\vskip2cm

\centerline{\bf Abstract}
\noi
The   recently
discovered $O(d,d)$ symmetry of the space of
   (slowly-varying) cosmological string vacua in $d+1$ dimensions
is shown to be preserved in the presence of
bulk string matter.
 The existence of $O(d,d)$ conserved currents allows all
 the equations of string
 cosmology to be reduced  to first-order differential equations.
The perfect-fluid approximation is not invariant
under $O(d,d)$ transformations,  implying that stringy
fluids possess in general a non-vanishing viscosity.

\vskip1.5cm

 \noindent
{CERN--TH.6321/91}\par
\noindent
{November 1991}\par
\vfill\eject
\bigskip
\eject

\footline={\hss\rm\folio\hss}
\pageno=1

{\bf 1. Introduction}

As a candidate theory for the unification of all interactions,
including gravity, string theory must possess all kinds of
 conventional
field-theoretic symmetries, in particular gauge and general
coordinate transformation invariance, as well as   their
supersymmetric generalizations.

 It has long been suspected that the above are only a tiny subset
of the full symmetries of string theory, and much research has gone
in recent years into trying to unravel the "higher" stringy symmetries
 which are not shared by conventional quantum field theories.

Although this program is still in its infancy, some interesting stringy
symmetries have indeed emerged, in particular those which go under the
generic name of target-space duality (or modular invariance) [1]. These
 symmetries might play a crucial role in connection with  compactification
 of the extra dimensions [2] and in determining the mechanism of
 supersymmetry breaking [3] in string theory.

It is very natural to ask whether some stringy
symmetries can also be found in the string analogue
 of the Einstein--Friedmann
equations, which govern  the time evolution of a spatially homogeneous
Universe, i.e. in what we shall refer to   as "string cosmology". Indeed,
it has been suspected for some time [4] that string cosmology should also
 exhibit a "duality" between large and small scale factors and/or large
and small temperatures. Finding the precise form  of this symmetry has
 proven to be a non-trivial task, however.

In the case of closed strings moving in a compact (target) space,
many authors [5,6] have recently discussed the generalization of
target-space duality to the non-static case. This leads to the
physical identification of  apparently unrelated cosmologies, in
the sense that the contraction of a compact dimension below the self-dual
 point is shown to be actually equivalent to its expansion above it.

It can be shown however that, even for open strings and/or a
 non-compact target space, duality-like discrete symmetries
 (scale-factor duality) survive [7] as symmetries of the field
equations that determine the possible consistent vacua of the theory.
In this case, more than of a symmetry, one should talk of a group acting
 on the space of solutions  by transforming non-equivalent
 vacua into each other.
 Furthermore, the discrete symmetry of the field equations persists [7]
 even in the presence of classical
string  sources.

An enlargement of the  symmetry group
 of cosmological string vacua in $d+1$ dimensions was discovered
 in Ref. [8]. Accordingly, discrete scale-factor duality is
 embedded in a continuous $O(d,d)$ "Narain" group [9]. Obviously,
 such a large group can only be interpreted as a symmetry of the
 equations of motion and not of the full theory. In more modern
 terminology, we would say that the generators of this group are
 "moduli" in the space of classical solutions.
  In this formulation the basic,
 $O(d,d)$-covariant objects are a shifted
dilaton $\Phi$, which absorbs the volume factor $\sqrt{|G|}$, and a
symmetric $O(d,d)$ matrix $M$, which mixes non-trivially the metric and
the antisymmetric tensor fields (thus implementing, in a perhaps
unexpected way,  Einstein's old dream of a non-symmetric unified
theory [10]).

Arguments for the  validity of the $O(d,d)$ symmetry
to all orders have been presented, both from the string
 field theory [11] and from the $\sigma$ model [12]
point of view. Extensions
 to the case of more general backgrounds that  are just independent
 of a subset of coordinates have also been given [13,12], and some amusing
 applications of the symmetry, both to cosmological
 solutions [14] and to $2D$ black holes [15] have been found.

In this paper we add classical string sources
to the manifestly $O(d,d)$ invariant (low-energy)
 cosmological vacuum equations of Ref. [12]. As in the case of
scale-factor duality, we find that manifest $O(d,d)$ invariance
is maintained in the presence of sources. This leads
 to the conclusion that
 string cosmology in $d+1$ dimensions is $O(d,d)$-covariant.
A welcome consequence of this symmetry is the existence of conserved
$O(d,d)$ currents. By constructing them explicitly we
are able to reduce all our constraints to first-order differential
equations. We shall also
derive a general continuity equation for the string sources, showing
that the antisymmetric field, unlike the dilaton,
 contributes explicitly to
the covariant conservation of the total source energy.

We stress that, in order not to spoil $O(d,d)$ invariance,
string sources must evolve in time in a way consistent with
the equations [16] describing the  motion of each string in the
(self-consistent) background  generated by the sources themselves.
 Thus, under $O(d,d)$, not only the backgrounds but also the sources
and their equations of state must change. Perhaps surprisingly, we shall
find that a perfect-gas equation of state in not invariant under $O(d,d)$,
 so that, generically, our cosmological backgrounds are sustained by
a string fluid with some specific kind of viscosity.
 \vskip 0.5 cm

{\bf 2. Background field equations with string sources}

At low energy, the tree-level effective action for closed string theory,
in $D$ dimensions, can be written as [17]
$$
I={1\over 2\kappa}\int d^D x \sqrt{|G|} e^{-\phi}
[R+(\nabla \phi)^2-V+{H^2\over 12}]\; , \eqno(2.1)
$$
where $H=dB$ is the antisymmetric tensor field strength,
$\kappa$ is a dimensionful parameter related to the string tension
 (see [7]), and $V$ is the cosmological constant
(proportional to $D-D^{crit}$). We shall consider
 in particular homogeneous
cosmological backgrounds which are independent of all space-like
coordinates, and for which a synchronous frame exists where
$G_{00}=-1$, $G_{0i}=0=B_{0i}$
($i,j=1,2,...,D-1 \equiv d$).

Defining [8]
$$
\Phi =\phi -  ln\sqrt{|detG|} \eqno(2.2)
$$
$$
M= \pmatrix{G^{-1} & -G^{-1}B \cr
BG^{-1} & G-BG^{-1}B \cr} \eqno(2.3)
$$
where $G$ and $B$ are $d \times d$ matrices representing respectively
$G_{ij}(t)$ and $B_{ij}(t)$, the action (2.1) can be written in a
manifestly $O(d,d)$-invariant form [8,12]:
$$
I=-{1\over 2\kappa}\int dt e^{-\Phi}[ \dot
 \Phi^2 +{1\over 8}Tr(\dot M \eta \dot M \eta)
+V]\; , \eqno(2.4)
$$
where $\eta$ is the $O(d,d)$ metric in off-diagonal form
$$
\eta = \pmatrix{0 & I \cr I & 0 \cr}\; ,
$$
a dot denotes differentiation with respect to the cosmic time $t$, and
we have allowed for a more
general scalar self-interaction $V(\Phi)$   than just a constant.

We also stress that $O(d,d)$-invariance takes a simple form
only if one works directly with the (unrescaled)
$\sigma$-model metric $G$. The fact
that $G$ is also the physical metric  of string theory
was shown in ref. [18] (see also [19]).

 Let us now add to the action (2.4) the contribution
of classical string sources. Their effective Lagrangian
(in the conformally flat gauge for the world-sheet metric)
can be written as
$$
L(t)={1\over 2\pi \a^\pr}\int d\sg d\tau \da (t-X^0(\sg,\tau))
(P_0{\pa_\tau X^0}+P_i{\pa_\tau X^i}-H) \; , \eqno(2.5)
$$
where
$$
H={1\over 2}[Z^T MZ-P_0^2-(X^{0 \pr})^2] \; ,\eqno(2.6)
$$
$P_0=-\pa_\tau X^0$ according to the string equations of motion, and
$$
Z^A(\sg,\tau)=(P_i,X^{\pr i}) \eqno(2.7)
$$
are the $2d$-dimensional phase-space coordinates
($\sg$ and $\tau$ are as usual the world-sheet variables, and a prime
denotes $\pa /\pa \sg$). Note that we write explicitly partial
differentiation with respect to $\tau$, as we have reserved a dot for
cosmic time derivatives. Moreover, a sum over different strings in
eqs. (2.5) and (2.6) is understood.

Since $\Phi$ is not directly coupled to the sources,
the variation of the total action with respect to $\Phi$ leads to the
same
dilaton equation as already obtained in [8], namely
$$
\dot \Phi^2 -2\ddot \Phi -
 {1\over 8} Tr[\dot M \eta \dot M \eta]+{\pa V\over \pa \Phi}-V=0 \; .
  \eqno(2.8)
$$
 The $G_{00}$ variation (see [8]) provides the (zero-energy) equation
$$
\dot \Phi^2 + {1\over 8} Tr[\dot M \eta \dot M \eta]-V=2\kappa
 \overline \r e^\Phi \; ,  \eqno(2.9)
$$
where
$$
\overline \r (t) \equiv \sqrt{|G|} \r =- {\da L\over \da G_{00}}
={1\over 4\pi \a ^\pr}\int d\sg {d\tau\over dX^0}
[P_0^2-(X^{\pr 0})^2] \eqno(2.10)
$$
represents the effective energy density of the string sources.

We have now to vary the action with respect to $M$. Proceeding as
in Ref. [12], we perform  an infinitesimal transformation
$$
\da M= \Om^T M\Om-M=\ep^TM+M\ep \; ,\eqno(2.11)
$$
where $\Om=I+\ep$ belongs to $O(d,d)$. One easily finds (using
$\eta^2=I$, $\eta \ep =-\ep^T \eta$) that the source contribution is
given by
$$
{\da L \over \da (\eta \ep)}=-{1\over 4\pi \a^\pr}\int
d\sg {d\tau\over dX^0}(ZZ^TM\eta-\eta MZZ^T)
\equiv -{1\over 2}(SM\eta-\eta MS) \; ,\eqno(2.12)
$$
where we have defined the symmetric matrix
$$
S(t)={1\over 2\pi \a^\pr}\int d\sg {d\tau \over dX^0} ZZ^T(\sg,\tau(\sg,
t)) \; . \eqno(2.13)
$$
By adding the massless field contributions (see [12]), the   variation
of the full action finally provides the equation of motion
$$
{d\over dt}(e^{-\Phi}M\eta \dot M)=2k \overline T \; , \eqno(2.14)
$$
where
$$
\overline T={1\over 2}(MS\eta -\eta SM) \eqno(2.15)
$$
represents a generalized "stress matrix" for the string sources.

The three equations (2.8), (2.9) and (2.14) are manifestly invariant
under the global transformation group defined by
$$
\Phi \ra \Phi~~~~,~~~~X^{\pr 0}\ra X^{\pr 0}~~~~,~~~~P_0 \ra P_0 \; ,
$$
$$
Z\ra \ti Z=\Om^{-1}Z~~~~,~~~~M \ra \ti M=\Om^T M \Om \; , \eqno(2.16)
$$
where $\Om$ is an $O(d,d)$ constant matrix satisfying
$$
\Om^T \eta \Om =\eta \; .  \eqno(2.17)
$$
It is important to stress that if a given set $\xi (\sg, \tau)=
\{  Z^A,P_0,X^{\pr 0} \}$ is a solution of the string equations in a
background $M$, than the transformed set $\ti \xi (\sg,\tau)=\{\ti
Z^A=\Om^{-1}Z, P_0, X^{\pr 0}\}$ is a solution for the transformed
background $\ti M$, as one can easily verify from the string equations
of motion following from the Hamiltonian (2.6):
$$
{\pa_\tau P_0}=-{\pa^2_\tau X^0}=-X^{0 \se}
-{1\over 2} Z^T\dot M Z \eqno(2.18)
$$
$$
{\pa_\tau Z}=(\eta MZ)^\pr \eqno(2.19)
$$
and from the constraints
$$
H=0 \eqno(2.20)
$$
$$
Z^T\eta Z+2P_0X^{\pr 0} =0 \; . \eqno(2.21)
$$
   The coupled string-background
system of equations (2.8), (2.9),
 (2.14), (2.18)--(2.21) defines for us string cosmology. When they
are all fulfilled, the stringy analogue of the stress tensor
  transforms covariantly under $O(d,d)$, namely
$$
\overline \r \ra \overline \r ~~~~,~~~~\overline T \ra \Om^T
\overline T \Om \; . \eqno(2.22)
$$

In view of future applications, it may be useful to express $\overline T$
in terms of the components of the usual energy--momentum tensor,
$\theta^{ij}$, and of the antisymmetric current $J^{ij}$ coupled to the
torsion field $B_{ij}$, which are defined by
$$
\overline \theta^{ij}=\sqrt{|G|}\theta^{ij}={\da L\over \da G_{ij}}
{}~~~,~~~~\overline J^{ij}=\sqrt{|G|} J^{ij}={\da L \over \da B_{ij}}\; .
\eqno(2.23)
$$
According to the Hamiltonian equations $\pa_\tau X^0=\pa H/\pa P_0$,
$\pa_\tau X^i=\pa H/\pa P_i$, one has
$$
P_0=-{\pa_\tau X^0}~~~,~~~ P_i={\pa_\tau X^j}G_{ji}
-X^{\pr j}B_{ji} \eqno(2.24)
$$
and by writing explicitly $L(t)$ in terms of $G$ and $B$, one
readily obtains
$$
\overline \theta^{ij}(t)={1\over 4\pi \a^\pr}\int d\sg {d\tau\over
dX^0}({\pa_\tau X^i}{\pa_\tau X^j}-
X^{\pr i}X^{\pr j}) \eqno(2.25)
$$
$$
\overline J^{ij}(t)={1\over 4\pi \a^\pr}\int d\sg {d\tau\over
dX^0}({\pa_\tau X^i}X^{\pr j}- X^{\pr i}
{\pa_\tau X^j}) \; . \eqno(2.26)
$$
{}From the definition (2.15) we are thus led to the explicit expression
$$
\overline T= \pmatrix {-\overline J, & -\overline \theta G +\overline J B
\cr
G\overline \theta -B\overline J, & G\overline J G +B\overline J B
-G\overline \theta B -B\overline \theta G \cr}
\eqno(2.27)
$$
where $\overline \theta$ and $\overline J$ represent the $d\times d$
matrices of eqs. (2.25) and (2.26).

We note, finally, that the conservation of the charge associated with the
global $O(d,d)$ invariance allows a first integration of eq. (2.14).
Let us define, indeed,
$$
\overline \Theta (t)= -{1\over 2} \int d\sg d\sg^\pr \ep (\sg -\sg^\pr)
F(\sg , \tau)Z(\sg,\tau)Z^T(\sg^\pr,\tau ^\pr)F(\sg^\pr,\tau^\pr)
 \eqno(2.28)
$$
$$
F=\eta-(X^{\pr 0}/\pa_\tau X^0)M \eqno(2.29)
$$
where $\ep (x)={1\over 2} sign(x)$ and $\tau (\sg,t)$
($\tau ^\pr(\sg^\pr,t)$) is solution of
$t=X^0(\sg,\tau)$ ($t=X^0(\sg^\pr,\tau^\pr)$). One finds:
$$
\dot {\overline \Theta} =-{1\over 2} \int d\sg d\sg^\pr \ep (\sg -
\sg^\pr)
\{[\pa_\tau X^0(\sg)]^{-1}[\pa_\tau F(\sg)Z(\sg)]Z^T(\sg^\pr)F(\sg^\pr)
$$
$$
+
F(\sg)Z(\sg)[\pa_\tau Z^T(\sg^\pr)F(\sg^\pr)][\pa_\tau X^0(\sg^\pr)]^{-1}
\}
$$
$$
 =-{1\over 2} \int d\sg d\sg^\pr \ep (\sg -\sg^\pr)
\{[\pa_\sg (\pa_\tau X^0)^{-1}MZ(\sg)]_t Z^T(\sg^\pr)F(\sg^\pr)
$$
$$
+
F(\sg)Z(\sg)[\pa_{\sg^\pr} Z^TM(\pa_\tau X^0)^{-1}(\sg^\pr)]_t
\} \eqno(2.30)
$$
where we used the equations of motion (2.19), and the following relation
between derivatives with respect to $\sg$   performed at constant $t$
and $\tau$:
$$
[\pa_\sg f]_t= [\pa_\sg f]_\tau - (X^{\pr 0}/\pa_\tau X^0)
[\pa_\tau f]_\sg
\equiv f^\pr -(X^{\pr 0}/\pa_\tau X^0)\pa_\tau f \; . \eqno(2.31)
$$
 Integrating finally  by parts in eq. (2.30) we get
$$
\dot {\overline \Theta} =\overline T \eqno(2.32)
$$
so that eq.(2.14) simply gives
$$
e^{-\Phi}M\eta \dot M=C(t)\equiv 2k \overline \Theta +A \; .
\eqno(2.33)
$$
Here $A$ is a constant antisymmetric matrix, and the matrix $C$, because
of the $O(d,d)$ properties of $M$ (i.e. $M\eta M=\eta$), satisfies the
property
$$
M\eta C=-C\eta M \; . \eqno(3.34)
$$
In the absence of sources ($Z=0, C=const$) one then finds the general
solution presented in [12].

In the next section we shall derive a first-order energy conservation
equation which, together with eqs. (2.9) and (2.33), implies also the
second order dilaton equation (2.8). We thus conclude that, thanks to
the $O(d,d)$ symmetry, the equations of string cosmology
 can always be reduced
to   first-order  differential equations.
\vskip 0.5 cm

{\bf 3. Covariant conservation of the source energy}

In general relativity, the energy--momentum tensor of the gravitational
sources is covariantly conserved as a consequence of the contracted
Bianchi identity. This identity could be applied to obtain a
conservation equation also in our case, of course, by re-writing the
field equations so as to include all the dilaton, torsion, and string
contributions to the "right-hand side" of a generalized Einstein
equation. However, such a generalized conservation law can be obtained
already in $O(d,d)$-invariant form, by using directly the
$O(d,d)$-covariant equations derived in the previous section.

Indeed, by differentiating eq. (2.9) with respect to cosmic time,
combining the result with eqs. (2.8) and (2.14), and by using the identity
$$
(M\eta \dot M\eta)^2=-(\dot M \eta)^2 \; , \eqno(3.1)
$$
we get the conservation equation in the form
$$
\dot {\overline \r} ={1\over 4}Tr[\overline T \eta M\eta \dot M \eta]
\eqno(3.2)
$$
or, equivalently
$$
\dot {\overline \r} ={1\over 4}Tr[S\dot M] \; .  \eqno(3.3)
$$

It is important to note that, according to this equation, there is no
direct contribution of the dilaton field to the covariant evolution of
the string-matter energy density (no dilaton-induced violation of the weak
equivalence principle). This is a usual result in many
scalar-tensor gravitational theories (like, for example, in Brans-Dicke
gravity), but a somewhat unexpected property in our context, where the
dilaton is directly coupled to the torsion part of the total Lagrangian,
and the general scalar-tensor theorems (see for instance [20]) are no
longer applicable.

We also note that eq. (3.3) is actually implied by the string
equations of motion. This is explicitly checked
by  writing, upon use of eq. (2.18)
$$
 {1\over 4}Tr[S\dot M]={1\over 8\pi \a^\pr}\int d\sg {d\tau \over dX^0}
Z^T\dot M Z = {1\over 4 \pi \a^\pr}\int d\sg {d\tau \over dX^0}
({\pa^2_\tau X^0}-X^{0 \se}) . \eqno(3.4)
$$
On the other hand, the explicit differentiation of eq. (2.10) gives
$$
\dot {\overline \r}(t)= {1\over 4\pi \a^\pr}
\int d\sg {d\tau \over dX^0}[\pa^2_\tau X^0-
\pa_\tau(X^{\pr 0})^2(\pa_\tau X^0)^{-1}] \; .\eqno (3.5)
$$
By using (2.31) one can easily see that the two integrands in eqs.(3.4)
and (3.5) differ by a total derivative in $\sg$ (at fixed $t$), thus
yielding eq. (3.3).

By working out explicitly the components of $\overline T$, the torsion
contribution to the conservation equation can be separated out as follows
$$
\dot {\overline \r} -{1\over 2}Tr[(\overline \theta G)(G^{-1}\dot G)]
+{1\over 2}Tr[\overline J \dot B]=0 \; . \eqno(3.6)
$$
For an isotropic, $D$-dimensional Friedmann--Robertson--Walker
(FRW) metric,
$G=a^2(t)I$, and, in the perfect fluid approximation ($\overline \theta
G=-\overline p I$, where $\overline p= \sqrt{|G|}p$ is the isotropic
pressure), eq. (3.6) takes the more familiar form
$$
\dot \r +(D-1)H(\r +p)+{1\over 2}Tr[J\dot B]=0 \; , H=\dot a/a ,\eqno(3.7)
$$
 which admits an interesting thermodynamical
interpretation.

Let us write indeed $\r=E/V$ and $J=\om/V$, where $E$ and $\om$ are,
respectively, the energy and the "torsional charge" of the source inside
a proper spatial volume $V=(a \ell)^d $ ($\ell=const$). Eq. (3.7) then
becomes, in differential form,
$$
dE+pdV=-{1\over 2}\om^{ij}dB_{ij} \; . \eqno(3.8)
$$
 This equation  suggests that, even if the source evolution is globally
adiabatic,
   entropy exchanges occur between the perfect-fluid
part and the "torsional" part of the source. In particular, a possible
damping of torsion in time, $\dot B <0$, should be accompanied by an
entropy increase in the fluid part.

 We note, finally, that
the thermodynamical role played by
$\om_{ij}$ in eq. (3.8) is
(formally) identical to that expected for the
intrinsic vorticity tensor,
in the context of a spinning-fluid model of the cosmological
sources [21].
\vskip 0.5 cm

{\bf 4. $O(d,d)$ transformations of the equation of state }

For the microscopic model of matter sources that we are considering,
based on classical strings, the source equation of state compatible with
a given background is determined by the solution of the string equations
of motion. It follows that, in the case of torsionless, isotropic FRW
backgrounds, sources with equation of state of the perfect-fluid type
are allowed, at least asymptotically, as discussed in previous papers
[16]. It should be stressed, however, that the presence of shear and
viscosity is in general required for a phenomenological fluid
description of the sources, even when the antisymmetric tensor is
vanishing. This point may be conveniently elucidated by recalling that,
in the context of our model, the matter sources transform in an
$O(d,d)$-covariant way, according to eq. (2.22).

Consider for example a perfect fluid, with given equation of state
($J=0$)
$$
\overline \theta ^i\,_j= -\overline p \da ^i_j~~~~,~~~~ p=\ga \r
\eqno(4.1)
$$
which is the source of a torsionless FRW background
$$
B=0~~~~,~~~~ G(t)=a^2(t) I \eqno(4.2)
$$
(we shall work, for simplicity, in $D=2+1$ dimensions, so that $I$ is
the $2\times 2$ unit matrix). We apply to $\overline T$ the
one-parameter $O(d,d)$ transformation with
$$
\Om (\a)={1\over 2}
\pmatrix {1+c & s & c-1 & -s \cr
-s & 1-c & -s & 1+c \cr
c-1 & s & 1+c & -s \cr
s & 1+c & s & 1-c \cr } \eqno(4.3)
$$
where $c=$cosh$\a$, $s=$sinh$\a$ (previously called "boost" [14], but
somewhat improperly since for $\a \ra 0$ it reduces not to the identity,
but to a matrix representing the discrete inversion of one scale
factor). For the transformed sources (denoted by a tilde) we then have
$$
\ti {\overline \r} =\overline \r ~~,~~ \ti {\overline J} =0 ~~,~~
\ti {\overline \theta}^i\,_j=-\overline p \tau^i\,_j \; ,
\eqno(4.4)
$$
where
$$
\tau = \pmatrix {c & -s \cr -s & -c \cr} \eqno(4.5)
$$

A perfect-fluid interpretation of this stress tensor is not possible, as
$\ti \theta$ is not diagonal. For a co-moving
viscous fluid, on the other hand,
the stress tensor can be written in general as [22]
$$
\theta ^i\,_j = -(p-\xi \vartheta )\da ^i\,_j +2\eta \sg^i\,_j \; ,
\eqno(4.6)
$$
where $\xi$ and $\eta$ are the bulk and shear viscosity coefficients,
$\vartheta=\nabla_\mu u^\mu$ ($u^\mu$
is the co-moving, geodesic velocity
field), and
$$
\sg^i\,_j=\nabla^i u_j -{\vartheta \over D-1} \da ^i\,_j \eqno(4.7)
$$
is the traceless shear tensor. For the metric obtained by applying to
$M$ the transformation (4.3):
$$
\ti G = {1\over 2ca^2}
\pmatrix {c(a^4+1)+a^4-1 & -s(a^4+1) \cr
-s(a^4+1) & c(a^4+1)-a^4+1 \cr} \; , \eqno(4.8)
$$
one finds that $\vartheta=0$, and
$$
\sg^i\,_j= \Ga_0\,^i\,_j =H \tau ^i\,_j \; . \eqno(4.9)
$$
A comparison with eq. (4.4) shows that the transformed sources can be
consistently described as a pressureless fluid with shear viscosity,
characterized by the equation of state
$$
\ti p =0 ~~~~,~~~~\ti \eta = -{\ga \ti \r \over 2H} \; . \eqno(4.10)
$$

Therefore, the perfect-fluid equation of state is not an
$O(d,d)$-invariant property, and viscosity is needed, in general,
for a phenomenological characterization of the sources.
Consistent equations of state (with viscosity) can be obtained by
applying directly $O(d,d)$ transformations to a known consistent
solution of the coupled string-background equations. An example worth
investigating could be, for $V=0$,
  the background
$$
\eqalign { B&=0~~,~~~\Phi =-{2\over D}(D-1)ln(\pm t/t_0) \cr
G&=a^2 I ~~~,~~~ a=(\pm t/t_0)^{\pm 2/D} \cr} \eqno(4.11)
$$
whose source is a perfect fluid with equation of state
$$
p=\pm {1\over (D-1)} \r \; . \eqno(4.12)
$$
Such an equation of state was shown [16] to be consistent
 with the string  equations of motion and constraints in the backgrounds
(4.11) at sufficiently small $t$.
\vskip 0.5 cm

We are grateful to P. Di Vecchia, K.A. Meissner and R. Pettorino
for useful discussions.

\vfill\eject
\centerline{\bf References}
\item{1.}K. Kikkawa and M. Yamasaki, Phys. Lett. B149 (1984) 357;

N. Sakai and I. Senda, Prog. Theor. Phys. 75 (1986) 692;

V. Nair, A. Shapere, A. Strominger and F. Wilczek, Nucl. Phys.

 B287 (1987) 402;

A. Giveon, E. Rabinovici and G. Veneziano, Nucl. Phys. B322 (1989) 167;

A. Shapere,  and F. Wilczek, Nucl. Phys. B320 (1989) 669.

\item{2.}G. Veneziano, Europhys. Lett. 2 (1986) 133;

T.R. Taylor and G. Veneziano, Phys. Lett. B212 (1988) 147.

\item{3.}A. Font, L.E. Iba{$\tilde n$}ez, D. L{$\ddot u$}st
 and F. Quevedo,
Phys. Lett. B245 (1990) 401;

S. Ferrara, N. Magnoli, T.R. Taylor and G. Veneziano, Phys. Lett. B245

 (1990) 409;

H.P. Nilles and M. Olechowski, Phys.Lett.B248(1990)268;

P. Binetruy and M.K. Gaillard, Phys. Lett. B216 (1991) 119.

\item{4.}Y. Leblanc, Phys. Rev. D38 (1988) 3087;

E.A. Alvarez and M.A.R. Osorio, Int. J. Theor. Phys. 28 (1989) 949;

R. Brandenberger and C. Vafa, Nucl. Phys. B316 (1989) 391.

\item{5.}P. Ginsparg and C. Vafa, Nucl. Phys. B289 (1987) 414;

T.H. Buscher, Phys. Lett. B194 (1987) 59;

T. Banks, M.Dine, H. Dijkstra and W. Fischler, Phys. Lett. B212 (1988) 45;

G. Horowitz and A. Steif, Phys. Lett. B250 (1990) 49;

E. Smith and J. Polchinski, Phys. Lett. B263 (1991) 59.

\item{6.}A.A. Tseytlin, Mod. Phys. Lett. A6 (1991) 1721;

A.A. Tseytlyn, in Proc.  First Int. A.D. Sakharov Conference on Physics,

 ed. by L.V. Keldysh et al. (Nova Science Pub., Commack, N.Y.,1991);

A.A. Tseytlin and C. Vafa, Harward Preprint HUTP-91/A049;

A.A. Tseytlin, Cambridge Preprint DAMPT-37-1991.

\item{7.}G. Veneziano, Phys. Lett. B265 (1991) 287.

\item{8.}K.A. Meissner and G. Veneziano, Phys. Lett. B267 (1991) 33
{}.
\item{9.}K.S. Narain, Phys. Lett. B169 (1986) 41;

K.S. Narain, M.H. Sarmadi and E. Witten, Nucl. Phys. B279 (1987) 369.

\item{10.}A. Einstein, Ann. Math. 46 (1945) 578.

\item{11.}A. Sen, Preprint TIFR/TH/91-35 (July 1991).

\item{12.}K.A. Meissner and G. Veneziano, Mod. Phys. Lett. A6 (1991) 3397.

\item{13.}A. Sen, Preprint TIFR/TH/91-37 (August 1991).

\item{14.}M. Gasperini, J. Maharana and G. Veneziano, Preprint
CERN-TH.6214/91 (to appear in Phys.Lett.B).

\item{15.}S.F. Hassan and A. Sen, Preprint TIFR/TH/91-40 (September 1991);

S. Khastgir and A. Kumar, Bubaneswar preprint (October 1991).

\item{16.}M. Gasperini, N. Sanchez and G. Veneziano,
Int. J. Mod. Phys. A6 (1991) 3853; Nucl. Phys. B364 (1991) 365.

\item{17.}C. Lovelace, Phys. Lett. B135 (1984) 75;

E.S. Fradkin and A.A. Tseytlin, Nucl. Phys. B261 (1985) 1;

C.G. Callan, D. Friedan, E.J. Martinec and M.J. Perry, Nucl. Phys. B262

 (1985) 593.

\item{18.}  N. Sanchez and G. Veneziano, Nucl. Phys. B333 (1990) 253.

 \item{19.} B.A. Campbell, A. Linde and K.A. Olive,
 Nucl. Phys. B355 (1991) 146.

\item{20.}J. Hwang, Class. Quantum Grav. 8 (1991) 1047.

\item{21.}J.R.Ray and L.L.Smalley, Phys.Rev.D27(1983)183;

M. Gasperini, Phys. Rev. Lett. 56 (1986) 2873;

A.J. Fennelly, J.C. Bradas and L.L. Smalley, Phys. Lett. A129 (1988) 195.

\item{22.}see for instance G.F.R. Ellis, in Proc. of the Int. School of
Physics "E. Fermi", Course XLVII (Varenna, 1965)
 (Academic Press, New York)
 p.104.

\end